\documentclass[twocolumn, times, showpacs,preprintnumbers,amsmath,amssymb,superscriptaddress, prl,reprint]{revtex4}

\usepackage{graphicx}

\usepackage{dcolumn}

\usepackage{bm}

\usepackage{color}

\DeclareGraphicsExtensions{.pdf,.png,.jpg,.eps}

\begin{document}

\title{Measurement of the 5D Level Polarizabilty in Laser Cooled Rb atoms}

\author{S.\,Snigirev$^*$}
 \affiliation{P.N. Lebedev Physical Institute, Leninsky prospekt 53, 119991 Moscow, Russia}
\affiliation{Russian Quantum Center, ul. Novaya 100, Skolkovo, Moscow region,
Russia}
\author{A.\,Golovizin}
 \affiliation{P.N. Lebedev Physical Institute, Leninsky prospekt 53, 119991 Moscow, Russia}
\affiliation{Russian Quantum Center, ul. Novaya 100, Skolkovo, Moscow region,
Russia}
 \affiliation{Moscow Institute of Physics and
Technology, 141704 Dolgoprudny, Moscow region, Russia}
\author{D.\,Tregubov}
 \affiliation{P.N. Lebedev Physical Institute, Leninsky prospekt 53, 119991
Moscow, Russia}
 \affiliation{Moscow Institute of Physics and
Technology, 141704 Dolgoprudny, Moscow region, Russia}
\author{S.\,Pyatchenkov}
 \affiliation{P.N. Lebedev Physical Institute, Leninsky prospekt 53, 119991
Moscow, Russia}
\author{D.\,Sukachev}
 \affiliation{P.N. Lebedev Physical Institute, Leninsky prospekt 53, 119991 Moscow, Russia}
\affiliation{Russian Quantum Center, ul. Novaya 100, Skolkovo, Moscow region,
Russia}
 \affiliation{Moscow Institute of Physics and
Technology, 141704 Dolgoprudny, Moscow region, Russia}
\author{A.\,Akimov}
 \affiliation{P.N. Lebedev Physical Institute, Leninsky prospekt 53, 119991 Moscow, Russia}
\affiliation{Russian Quantum Center, ul. Novaya 100, Skolkovo, Moscow region,
Russia}
\author{V.\,Sorokin}
 \affiliation{P.N. Lebedev Physical Institute, Leninsky prospekt 53, 119991 Moscow, Russia}
\affiliation{Russian Quantum Center, ul. Novaya 100, Skolkovo, Moscow region,
Russia}
\author{N.\,Kolachevsky}
 \affiliation{P.N. Lebedev Physical Institute, Leninsky prospekt 53, 119991 Moscow, Russia}
\affiliation{Russian Quantum Center, ul. Novaya 100, Skolkovo, Moscow region,
Russia}
 \affiliation{Moscow Institute of Physics and
Technology, 141704 Dolgoprudny, Moscow region, Russia}




$^*$E-mail: ss@rqc.ru

\begin{abstract}
We report on accurate measurements of the scalar $\alpha_S$ and
tensor $\alpha_T$ polarizabilities of the 5D fine structure levels
5D$_{3/2}$ and 5D$_{5/2}$ in Rb. The measured values (in
atomic units) $\alpha_S(\textrm{5D}_{3/2})=18400(75)$,
$\alpha_T(\textrm{5D}_{3/2})=-750(30)$,
$\alpha_S(\textrm{5D}_{5/2})=18600(76)$ and
$\alpha_T(\textrm{5D}_{5/2})=-1440(60)$ show reasonable
correspondence to previously published theoretical predictions,
but are more accurate. We implemented laser excitation of the 5D
level in a laser cooled cloud of optically polarized  Rb-87 atoms
placed in a constant electric field.

\end{abstract}

\maketitle

\section{Introduction}

Study of   atomic and molecular polarizabilities  remains an
important task in atomic physics. The atomic polarizability

\begin{equation} \label{eq0}
\alpha_\gamma=\sum_{\gamma'}\frac{|\langle\psi_\gamma|e\bm{r}|\psi_{\gamma'}\rangle|^2}{E_\gamma-E_{\gamma'}}
\end{equation}
depends on  electric dipole matrix elements
$|\langle\psi_\gamma|e\bm{r}|\psi_{\gamma'}\rangle|^2$ \cite{Landau}
which also describe transition strengths, state lifetimes, van der
Waals interactions,
 and scattering cross-sections. Here $e\bm{r}$ denotes an electric dipole
 operator,  $E_\gamma$ the level energy with quantum
 number $\gamma$, and $\psi_\gamma$ its wave functions.
 Accurate measurements of
  polarizability facilitate progress in sophisticated atomic structure calculations and
  the theory of heavy atoms which results in more precise predictions for other  important atomic
 parameters (see e.g. \cite{Mitroy}).

Measurements of polarizabilities become even more crucial
in applications for modern optical atomic clocks. Predictions of
the ``magic'' wavelength in optical lattice clocks \cite{Katori}
and accurate estimation of the  blackbody radiation shift require
precise knowledge of static and dynamic polarizabilities
\cite{Derevianko}. Measurement of static polarizabilities provides
an important benchmark for calculations resulting in significant
improvement of optical clock performance \cite{PTB, Oates}. No
less important are polarizability measurements for the ground
state hyperfine components of the alkali atoms used in microwave atomic clocks (see, e.g., \cite{Weiss}).

For alkalis in the ground state the uncertainty in the theoretical prediction for the
polarizability is about 0.1\% \cite{Babb} while the measurement
uncertainty is typically 0.5\,-\,1.0\% (see \cite{Vigue,Gould}).
The lowest uncertainty is demonstrated by using laser cooled atoms
and atomic interferometers providing high sensitivity to electric
fields \cite{Hollmergen}. Ground state atoms are relatively easy
to prepare in a particular hyperfine and magnetic quantum state
while the natural decay does not pose any limitation for the
experiment. 

On the other hand, relatively long-lived Rydberg
atoms are highly sensitive to electric fields  \cite{Pfau} which
simplifies interpretation of the experimental results.
Polarizability measurements were performed in atomic vapor cells
\cite{Sullivan} and on laser cooled atoms \cite{JIA} with relative
uncertainties of  0.1-3\% depending on the state. Asymptotic theory
of Rydberg atoms is well understood and shows good agreement with
experimental observations.

However, atoms in intermediately excited states pose a challenge both for experiment and theory. They are typically
short-lived and difficult to address, while the response to an
electric field is  small compared to the Rydberg states. For
example,  the intermediate states in Rb and Cs ($n=6-10$) were
studied previously using atomic beams  (see, e.g.,
\cite{Svanberg}). In the cited reference a scalar polarizability
was measured with a relative uncertainty of about 5\%.
Calculations of these states are also less accurate  since the sum
(\ref{eq0})  contains terms of alternating signs cancelling each
other while a numerical error accumulates.

In this paper we report an accurate measurement of the  static
scalar and tensor polarizabilities of the $5\textrm{D}_{3/2}$ and
$5\textrm{D}_{5/2}$ levels in Rb-87 using spectroscopy of
laser cooled atoms in a dc electric field. To our knowledge, the
polarizability of the 5D level in Rb has not been measured to date.

The 5D level in Rb is used in  metrology \cite{Nez,Tetu}
because the frequency of the 5S-5D transition is recommended by
the International Committee for Weights and Measures (CIPM) for
the practical realization of the definition of the meter
\cite{Quinn}. Knowledge of the 5D level polarizability is
essential for an accurate evaluation of systematic shifts.

However, published calculations  show considerable
discrepancy. Two approaches were
 implemented  to calculate the polarizabilities of the 5D
level in Rb: the method of model potential
\cite{Manakov,Ovsiannikov} and the regular second order
perturbation theory with direct summation of matrix elements and
 integration over the continuous spectrum \cite{Beigman}. In the
 latter case the transition probabilities were calculated by the
 program ATOM \cite{ATOM} partly relying on an accurate experimental
 input. The calculated results  \cite{Ovsiannikov} and \cite{Beigman}
 differ 30\% in the scalar
 polarizability and more than 100\% in its tensor component as shown in Table \ref{table1}.
 Although this discrepancy can be readily explained by the intrinsic
 uncertainty of the theoretical approach \cite{Ovsprivcomm}, an
 accurate experimental measurement of the polarizability
 components is highly desirable.

\begin{table} [t!]\caption{Calculated values of the scalar $\alpha_S$ and tensor $\alpha_T$
polarizabilities of the 5D$_{3/2}$ and 5D$_{5/2}$ fine structure
levels in Rb atoms according to \cite{Ovsiannikov, Beigman}. The
values are given in atomic units $a_0^3$, where $a_0$ is the Bohr
radius. The discrepancy is due to different theoretical
approaches. }\label{table1}
\begin{ruledtabular}
\begin{tabular} {c|c|c|c|c} 
  Ref. & $\alpha_S(5\textrm{D}_{3/2})$  & $\alpha_T(5\textrm{D}_{3/2})$ & $\alpha_S(5\textrm{D}_{5/2})$&$\alpha_T(5\textrm{D}_{5/2})$ \\
 \hline
 \cite{Ovsiannikov}&21 110&-2871&20 670 &-3387\\
\cite{Beigman}&16 600&-1060&16 200 &-909 \\
\end{tabular}
\end{ruledtabular}
\end{table}

 Using laser cooled Rb atoms placed
 in the center of a plane capacitor we managed to reach a relative
 uncertainty for the scalar polarizability of 0.4\% which is
 comparable to measurements in the ground state. Optical pumping of atoms to a certain magnetic sublevel allowed us to measure
 the tensor polarizability component with an uncertainty of 4\%.
 The measured values allow for distinction between the
 results of calculations
 and may facilitate further theoretical progress.

\section{The Stark effect on 5P$_{3/2}$- 5D$_{3/2,\,5/2}$ transition}

If an atom is placed in an external electric field, it becomes
polarized and its energy levels are shifted according to
\cite{Landau}:

\begin{equation}\label{eq1}
\Delta E = -\frac{1}{2}( \alpha_S  +  \alpha_T
 P)\mathcal{E}_z^2\,.
\end{equation}

Here  $\alpha_S$ and $\alpha_T$ are the scalar and tensor
polarizabilities, respectively, while for alkali atoms the parameter $P$
can be written as:

\begin{equation}\label{eq2}
P= \frac{[3 m_F^2 -
F(F+1)][3Q(Q-1)-4F(F+1)J(J+1)]}{(2F+3)(2F+2)F(2F-1)J(2J-1)}\,
\end{equation}

with  $Q=F(F+1)+J(J+1)-I(I+1)$. Here $m_F$ is the magnetic quantum
number, and $F$, $J$, $I$    are the total magnetic moment, the
electron magnetic moment and the nuclear spin  quantum numbers,
respectively. The tensor component describes the relative
splitting of magnetic sublevels in the multiplet and equals 0 for
states with $J=0$ and $J=1/2$. To measure both scalar and tensor
polarizabilities one should control the atomic state and
address different magnetic and hyperfine sublevels.

If laser spectroscopy is used to probe the Stark effect, both
ground  and excited  levels are  shifted in the external electric
field. In that case the resonance frequency will be shifted according to
\begin{equation}\label{eq3}
\Delta f = -\frac{1}{2h} \big(\alpha_{S}(e)-\alpha_{S}(g)  +
 \alpha_{T}(e) P_e-\alpha_{T}(g) P_g \big) \mathcal{E}_z^2\,,
\end{equation}
where  $g$ and $e$ stand for the ground and excited states,
respectively, and $\Delta f$ is the shift of the resonance
frequency.

For non-degenerate states, the contributions of the individual
transitions between the magnetic sublevels is proportional to the
relative probabilities $|\langle F_g,\,m_{F_g}|e\bm{r}|F_e,\,m_{F_e}\rangle|^2$ according to \cite{Steik}:
\begin{eqnarray}\label{eq4}
|\langle F_g,\,m_{F_g}|e\bm{r}|F_e,\,m_{F_e}\rangle|^2=\\
\nonumber %
=|\langle J_g|e\bm{r}|J_e\rangle|^2(2F_e+1)(2J_g+1)(2F_g+1)\times\\
\nonumber %
\times \left\{%
\begin{array}{ccc}
  J_g & J_e & 1 \\
  F_e & F_g & 1 \\
\end{array}%
\right\}^2
\left(%
\begin{array}{ccc}
  F_e & 1 & F_g \\
  m_{F_e} & q & -m_{F_g} \\
\end{array}%
\right)^2\,,
\end{eqnarray}

where  $q=0$ for $\pi$ polarized light and $q=\pm1$ for
$\sigma^{\pm}$, and the matrices are 6-$j$ and 3-$j$ symbols,
respectively. This relation should be taken into account if
multiple magnetic sublevels are populated and the corresponding
spectral components are not well resolved.

In our case the ground state is the 5P$_{3/2}$ level in Rb and the
excited state is the 5D level, which are coupled by 776\,nm laser
radiation. The experimental values for scalar and tensor
polarizabilities of the  5P$_{3/2}$ level are equal to
$\alpha_S(\textrm{P}_{3/2})=859(7)a_0^3$ and
$\alpha_T(\textrm{P}_{3/2})=-163(3)a_0^3$ \cite{Windholz}.
 The atomic unit of the polarizability is the cube of the
Bohr radius $a_0^3=1.4818\times 10^{-31}\ \textrm{m}^3$, but in
the experiment the units of $\textrm{Hz(V/cm)}^{-2}$ are more
practical. The conversion is given by
$\alpha[\textrm{Hz}\textrm{(V/cm)}^{-2}]=2.482\times10^{-4}
\alpha[a_0^3]$.

\section{Experiment}

\begin{figure}
\includegraphics[width=0.4 \textwidth]{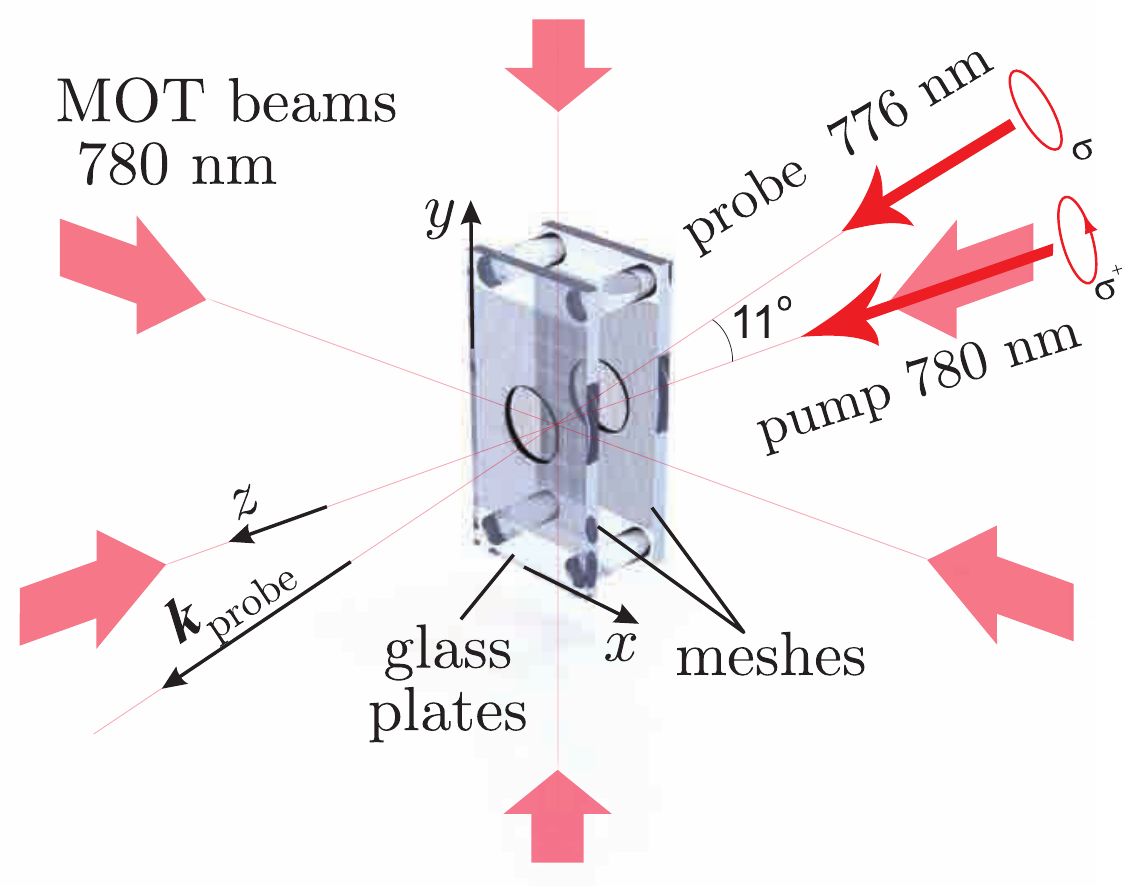}
\caption{ (Color online) Capacitor
assembly and laser beam orientation. The MOT is formed close to the
center of a flat capacitor consisting of two 80\% transparent
meshes glued to parallel glass plates with a mesh separation of 1\,cm.
The circularly $\sigma^+$ polarized pump beam at 780\,nm is
parallel to the electric field ($z$-axis) while the probe beam is
directed at the angle of $11^\circ$ in the $z$-$y$ plane to the pump beam and is also
circularly polarized (either $\sigma^+$ or $\sigma^-$) with respect
to its wave vector $\bm{k}_\textrm{probe}$. } \label{fig1}
\end{figure}

To measure the Stark shift of the 5D level in Rubidium we
used two-stage laser excitation $5S\rightarrow5P\rightarrow5D$ in
an external dc electric field on the order of  1\,kV/cm.  Rb-87
atoms were laser cooled in a regular six-beam magneto-optical trap
(MOT) with an axial magnetic field gradient of up to 20\,G/cm. The
cloud of 300\,$\mu$m in diameter contains about $10^6$ atoms at a
temperature of 300\,$\mu$K. The MOT configuration  and the
 excitation scheme are similar to
one described in Ref \cite{Snigirev}. Compared to
Ref.\,\cite{Snigirev}, the atomic cloud was formed in the center of
a plane capacitor consisting of two metallic meshes, as shown in
Fig.\ref{fig1}. The capacitor was placed inside a vacuum glass
cell (3\,cm$\times$3\,cm$\times$12\,cm) providing easy optical
access.

\begin{figure} [t!]
\includegraphics[width=0.95 \linewidth]{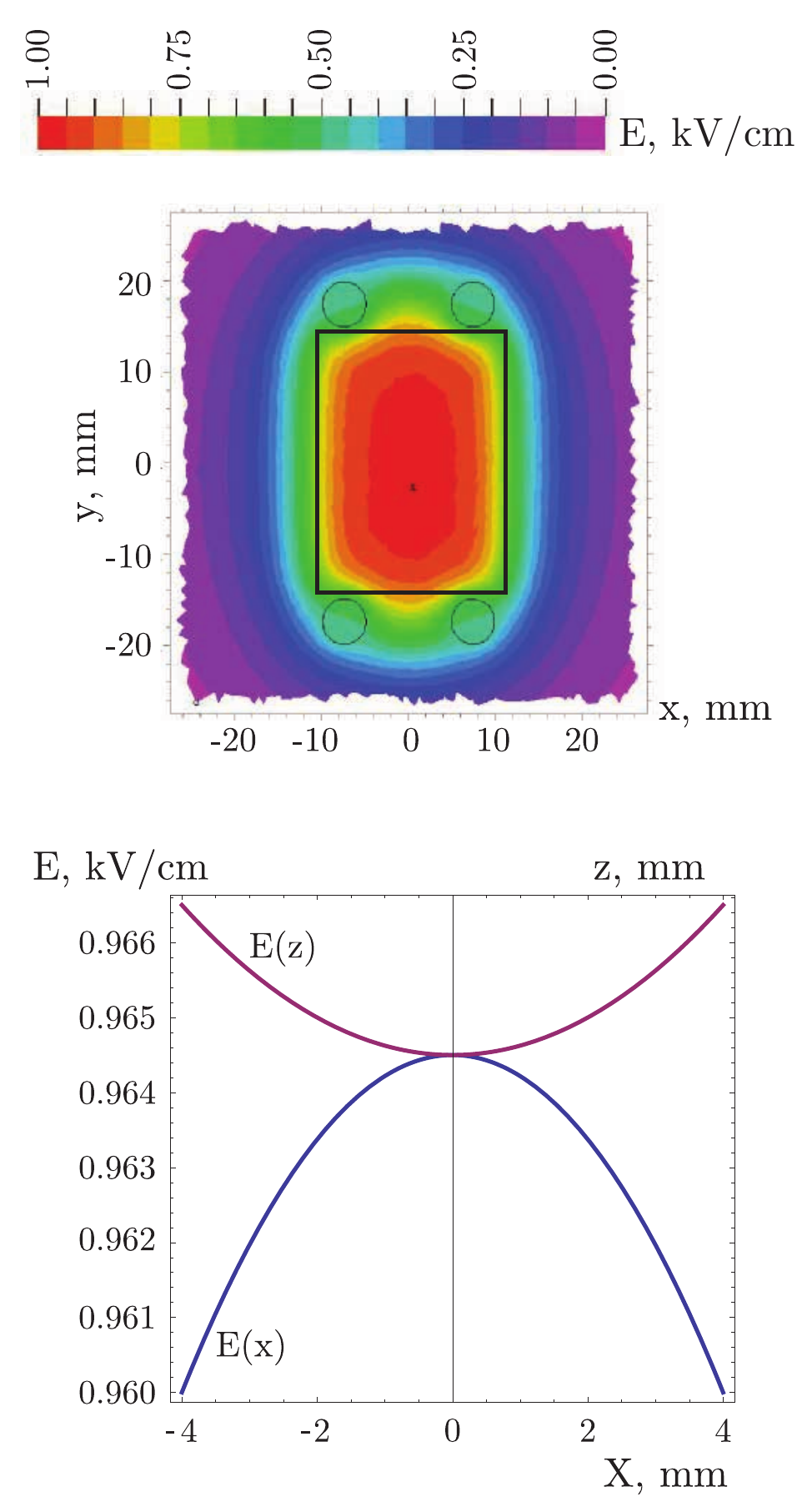}
\caption{  (Color online) Numerical calculations of the electric field
distribution with 1 kV applied to the capacitor depicted in  Fig.\ref{fig1}. Top:
field distribution at the central plane ($x$-$y$) of the
capacitor. \\ Bottom: zoom in of the central volume. The origin is
positioned at the geometric center of the capacitor.  }
\label{fig2}
\end{figure}

The mesh consists of non-magnetic stainless steel wires with a diameter of
25\,$\mu$m and has an optical transparency of 80\%. To
manufacture one of the capacitor plates the mesh was made taut and
then glued to a flat glass plate with a hole of 1\,cm in diameter.
We saw to it that glue did not penetrate through the mesh to
the front surface. Two plates were glued together using four glass
posts of calibrated length, thus forming a plane capacitor with
rectangular plates of $2\,\textrm{cm}\times 3$\,cm and the
separation of 1\,cm. The distance between the plates and the hole size provides clearance for the laser cooling beams.
One pair of laser cooling beams were sent through the mesh as
shown in Fig.\ref{fig1}, which did not significantly change the
cloud shape or the number of atoms.

Although all glass components were manufactured in the Laboratory of
Optics, P.N. Lebedev Physics Institute, and have superior flatness
and well-defined sizes within a few $\mu$m, the glue can influence the
distance $l$ between the meshes. To reach an uncertainty of
0.5\% in polarizability one should know the electric field to
0.2\%, which corresponds to 20\,$\mu$m uncertainty in the distance.

The parallelism of the glass plates was checked in the air using a
micrometer and found to be parallel to within
$15^\prime$. This angle was taken into account in calculations of the electric field. The distance $l$ between the meshes was measured
optically in vacuum  using a high NA lens assembly ($F/D\approx1$ ) which
imaged a free standing mesh  on a CCD camera. The lens and the
camera were rigidly placed on a three-coordinate translation
stage outside the vacuum chamber. The translation stage axis was
 aligned with $1^\circ$ accuracy perpendicular to the
capacitor plate (along the $z$-axis, Fig.\,\ref{fig1}). The focus
position for one of the meshes was determined from a number of
shots using a gradient filter method and then by fitting the position
of the translation stage to the highest image sharpness. This method
provides a statistical uncertainty of 5\,$\mu$m. Moving the
translation stage in the $x$-$y$ direction we performed similar
measurements in the region within $\pm$3\,mm of the hole center.
The distance remained constant within 20\,$\mu$m, the scatter
can be explained by the mesh thickness. The final result
including the averaged mesh thickness gives
$l=9.88(3)\,\textrm{mm}$. The refractive index of air contributes to
the result on a negligible level.

The field distribution in our capacitor was simulated using a finite element analysis, the result for the $x$-$y$ plane as well as the zoom in
of the central volume is shown in Fig.\ref{fig2}(a).  The position of
the atomic cloud was controlled within  $\pm1$\,mm with respect to
the center of the capacitor by two CCD cameras. As follows
from Fig.\,\ref{fig2}, the  field variation within this
volume is
 less than 0.05\%. The potential difference between the plates
could be varied from zero to 2.5\,kV using a high-voltage power
supply (Stanford Research Systems PS350). The accuracy of the device was studied with a high-precision voltmeter and
corresponded to 0.1\,\% variation.

\begin{figure}
\includegraphics[width= \linewidth]{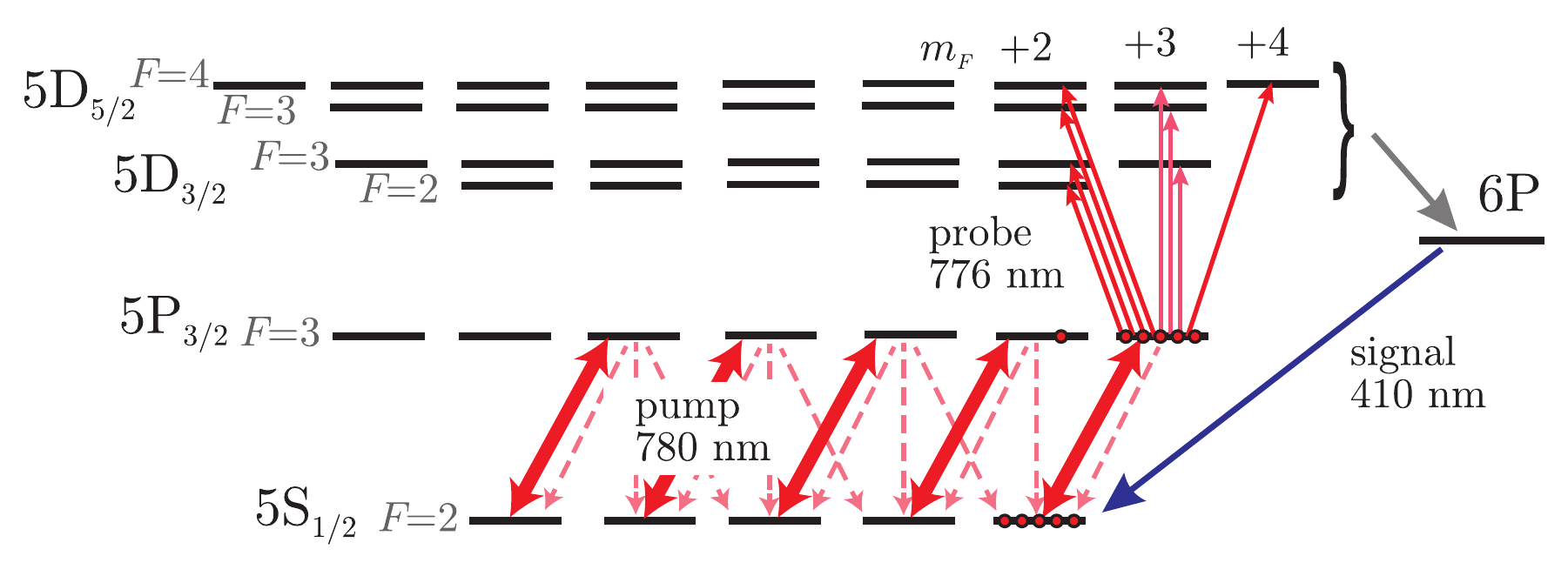}
\caption{ (Color online) Relevant energy levels in Rb-87 (not to scale) and
excitation laser fields. Radiation at 780\,nm coupling 5S$_{1/2}$
and 5P$_{3/2}$ levels was used for laser cooling and for optical
pumping of Rb atoms. The pump beam was  $\sigma^+$ polarized and
directed along the $z$-axis (see Fig.\,\ref{fig1}), transferring
population to the $5\textrm{P}_{3/2} (F=3, m_F=+3)$ level. The
polarized probe laser beam at 776\,nm  was tuned in resonance to
one of the hyperfine transitions
$5\textrm{P}_{3/2}\rightarrow5\textrm{D}_{3/2,\,5/2}$.  Excitation
of the 5D level was measured by spontaneous emission of 410\,nm
photons via the 6P level.
 } \label{fig3}
\end{figure}

\begin{figure}
\includegraphics[width= 0.95\linewidth]{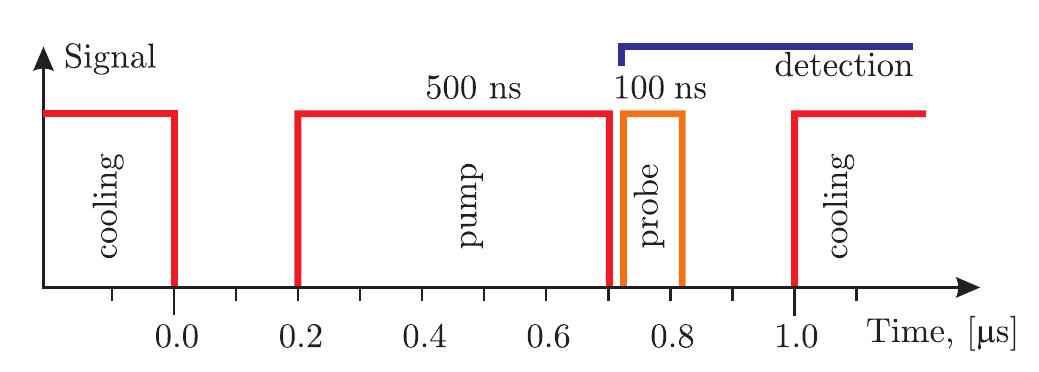}
\caption{(Color online) Pulse sequence in our experiment. After laser cooling for
 19\,$\mu$s the cooling laser was switched off for 1\,$\mu$s  for optical
 pumping (780 nm, 500\,ns) and probing the 5P$\rightarrow$5D transition
 (776 nm, 50\,ns). To avoid the ac Stark shift from  strong 780\,nm pumping
 radiation while obtaining enough signal, the probe pulse was switched on 10\,ns
  after the pump was switched off.  Detection starts simultantously with the probe beam and lasts for $1\,\mu$s. The whole cycle was repeated every
 20\,$\mu$s, the MOT magnetic field was continuously on.
 } \label{fig4}
\end{figure}

Measurement of the Stark shift of  each of the
$5\textrm{P}_{3/2}\rightarrow5\textrm{D}_{3/2,\,5/2}$ transitions
(Fig.\,\ref{fig3}) was performed in a pulsed regime, the pulse sequence
was repeated every 20\,$\mu$s and is depicted in Fig.\,\ref{fig4}.
First, atoms were laser cooled using the 780\,nm transition coupling
$5\textrm{S}_{1/2}$ and $5\textrm{P}_{3/2}$ levels. Since the MOT
is formed close to the zero of the quadrupole magnetic field, all
magnetic sublevels of the $5\textrm{P}_{3/2}$ level become nearly
equally populated, which does not allow for determination of
polarizabilities (\ref{eq1}), so we prepare the atoms into a particular
magnetic state using optical pumping at 780\,nm.

After switching off the MOT beams we waited for 200\,ns and
applied a circularly $\sigma^+$ polarized pump pulse along the
$z$-axis (Fig.\,\ref{fig1}) to transfer atoms to the
$5\textrm{P}_{3/2} (F=3, m_F=+3)$  magnetic state, as shown in
Fig.\,\ref{fig3}. The pulse had an intensity of  100 mW/cm$^2$ and
a duration of 500\,ns, which is much longer than the reverse
pumping rate. Experimental results presented in the next Section
verify that nearly all atoms  addressed by the probe
5P$\rightarrow$5D radiation were pumped into the $m_F=+3$
sublevel.

The probe beam at 776\,nm was tuned close to resonance with
one of the fine structure sublevels 5D$_{5/2,\,3/2}$, which are separated by 89\,GHz.
The probe beam was directed at an angle of 11$^\circ$ to the
$z$-axis, allowing for independent control of its polarization. Polarization of the probe beam was changed from
$\sigma^+$ to $\sigma^-$ with respect to its wave vector
$\bm{k}_\textrm{probe}$. In this case we can address different magnetic
sublevels of the 5D multiplet and thus derive polarizabilities
using (\ref{eq1}). Due to the set angle to the quantization axis, probe radiation always contains an admixture of linear
polarized light (with respect to the $z$-axis), which is taken into
account in our analysis. Thus, with the pump beam we couple the
sublevel $5\textrm{P}_{3/2} (F=3, m_F=+3)$ either with one of the
sublevels from $5\textrm{D}_{5/2}(F=3,\,4,\,m_F=+2,...,F)$
multiplet or with one from the
$5\textrm{D}_{3/2}(F=2,\,3,\,m_F=+2,...,F),$  as shown in
Fig.\,\ref{fig3}. The hyperfine structure of the 5D level is about 100\,MHz and is well resolvable.

The probe pulse with an intensity of 100 mW/cm$^2$ was switched on
right after the pump beam was switched off. The time delay between
the two pulses was chosen to be 10\,ns to avoid overlap between pulses.
The strong pump beam causes an ac Stark shift of the 5P$_{3/2}$ level,
which influences the results of our measurement. The 5P$_{3/2}$ level
lifetime  equals 30\,ns and most of the atoms excited to the
5P$_{3/2}$  remain there when the probe beam is on. The probe
pulse duration was 50\,ns, which is much shorter than the
5D level lifetime (300\,ns). This prevents optical pumping back to
the 5P$_{3/2}$ level and
 re-distribution of the population between magnetic
components.

Approximately 30\% of atoms in the 5D state decay to the
ground state {\it{via}} the 6P level, emitting 410\,nm photons. In
our experiment, "blue" photons were collected onto a photomultiplier
tube equipped with a narrow-band 410\,nm filter. Photons were
counted in a time window of 1 $\mu$s, as shown in Fig.\,\ref{fig4}.
The probe laser was scanned over the resonance with  an
acousto-optical modulator (AOM).  For each of the frequency steps
(typically 50 per line)
 the signal was accumulated for 0.1\,s. A typical count rate at
 the resonance position was $2\times 10^4$cps.

The MOT magnetic field gradient was  continuously switched on.
The influence of magnetic field on our results is discussed below.

\begin{figure}[t!]
\includegraphics[width= 0.9\linewidth]{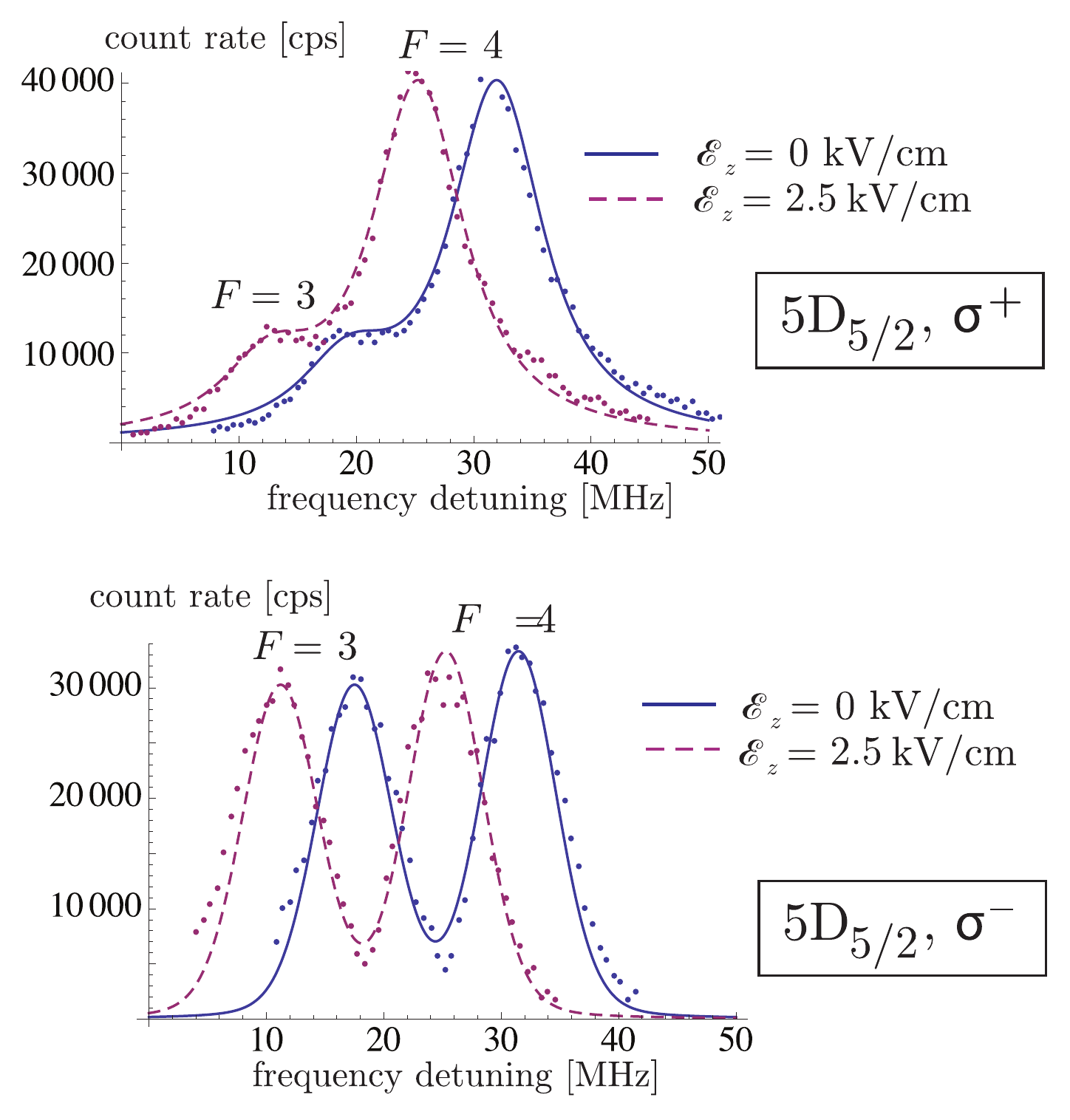}
\caption{(Color online) Top:
spectral line profiles of the
$5\textrm{P}_{3/2}\rightarrow5\textrm{D}_{5/2}$ transition with
the  probe beam $\sigma^+$ polarized along
$\bm{k}_\textrm{probe}$. The dashed line corresponds to zero electric
field, $\mathcal{E}=0$\,kV/cm, and the solid line to
$\mathcal{E}=2.5$\,kV/cm. The hyperfine component $F=3$ is excited
because the probe beam possesses a fraction  of linear polarization
if projected onto the $z$-axis. Bottom: the same experiment, but for
the probe beam $\sigma^-$ polarized along its wave
vector. The amplitude of $F=3$ significantly increases because the
transition to the sublevel $5\textrm{D}(F=3,\,m_F=2)$ becomes
allowed (see also Fig.\,\ref{fig3}). Fits are done by the model
function (c) described in the text.
 } \label{fig5}
\end{figure}

Measurement of the Stark shift for each of the lines
$5\textrm{P}_{3/2}\rightarrow5\textrm{D}_{3/2,\,5/2}$ was
performed at different voltages on the capacitor plates. For each
of the selected voltages three spectral lines were recorded, namely,
with positive voltage polarity, grounded electrodes and negative polarity,
to prevent any influence of  charging. The zero-voltage data was also
used in the data  analysis.

\begin{figure}[t!]
\includegraphics[width= 0.9\linewidth]{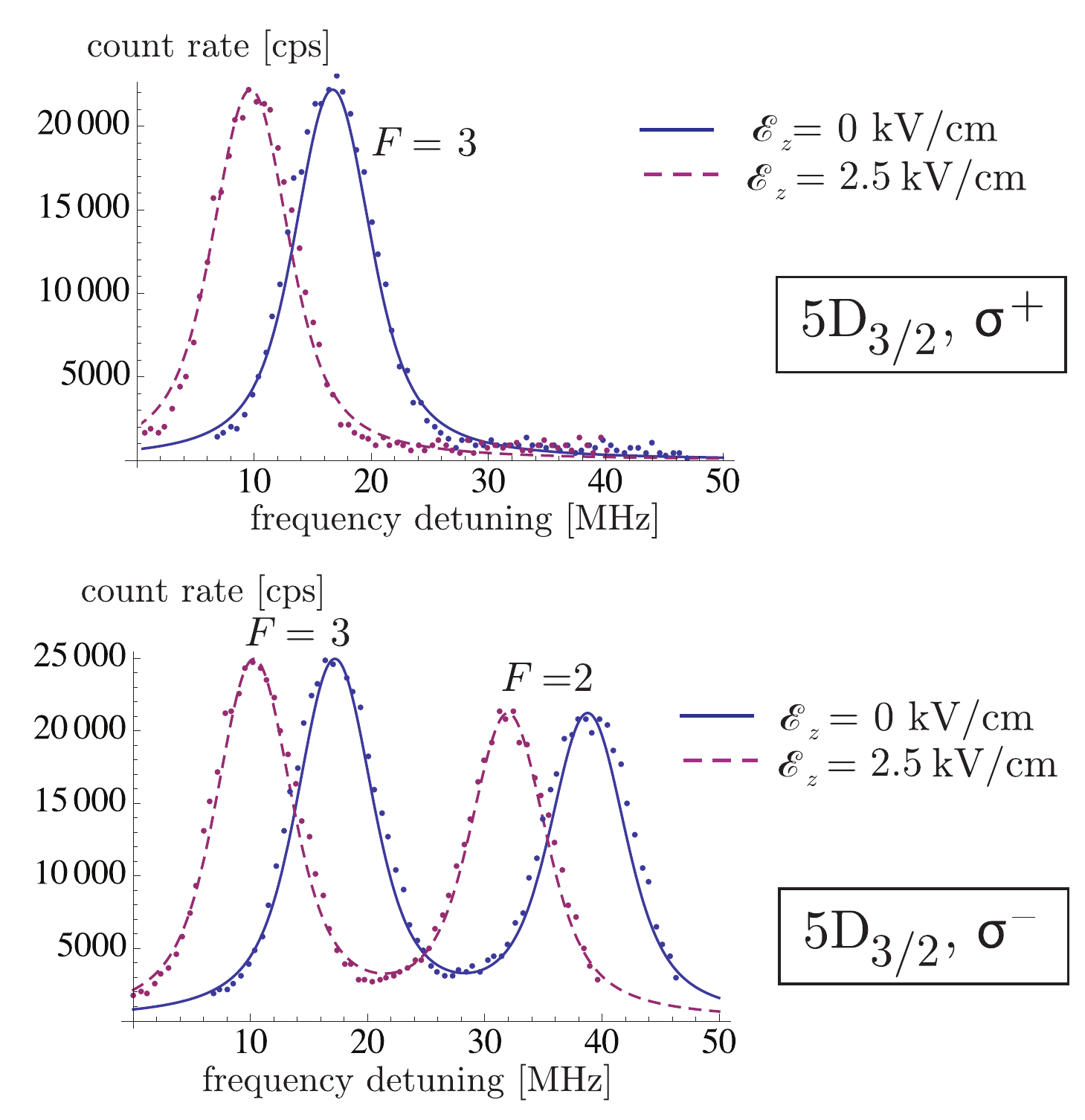}
\caption {(Color online) Top: spectral line
profiles of the $5\textrm{P}_{3/2}\rightarrow5\textrm{D}_{3/2}$
transition with the probe beam $\sigma^+$ polarized
with respect to $\bm{k}_\textrm{probe}$. The dashed line corresponds to
zero electric field, $\mathcal{E}=0$\,kV/cm, and the solid line to
$\mathcal{E}=2.5$\,kV/cm. The hyperfine component $F=3$ is clearly
visible due to the component of linear polarization, but  $F=2$ is not distinguishable from the noise. The spectrum
proves the high efficiency of optical pumping and the absence of
$\sigma^-$ component.
 Bottom: same experiment, but for the probe beam $\sigma^-$
polarized with respect to $\bm{k}_\textrm{probe}$. Now the component
with $F=2$ is efficiently excited according to Fig.\,\ref{fig3}.
Fits are done by the model function (c) described in the text.
 } \label{fig6}
\end{figure}

\section{Results and Data Analysis}

Typical spectra of $5\textrm{P}_{3/2}\rightarrow5\textrm{D}_{5/2}$
and $5\textrm{P}_{3/2}\rightarrow5\textrm{D}_{3/2}$ transitions at
different electric field strengths and probe beam polarizations
are shown in Figs.\,\ref{fig5},\,\ref{fig6}, respectively. From
these spectra we derive the frequency shifts of the corresponding
transitions, estimate the quality of initial state preparation and
estimate the excitation probabilities to different magnetic
sublevels of the 5D level.

The spectra were fitted by three different models:\ (a) sum of two
Gaussian functions, (b) sum of two Lorentzian  functions and (c)
sum of  two Lorentzian functions convoluted with the probe pulse
spectral profile. The  pulse shape was
measured in \cite{Snigirev} and its Fourier spectrum
width equals 20 MHz for 50\,ns duration. The model used 6 fit
parameters: two amplitudes, two central frequencies and two widths. The different fit
functions were used to test the model dependency of our fitting
procedure and to evaluate the corresponding uncertainty. The fit
gave us the frequency of each of the hyperfine components at
different values of the electric field.

To derive the polarizability of the $5\textrm{D}_{5/2}$ level we used
transitions from the $5\textrm{P}_{3/2}(F=3)$ level to
two different hyperfine sublevels excited at different
polarizations of the probe beam: $5\textrm{D}_{5/2}(F=4)$
(Fig.\,\ref{fig5}, top) and $ 5\textrm{D}_{5/2}(F=3)$
(Fig.\,\ref{fig5}, bottom). For the $5\textrm{D}_{3/2}$ level we
use, respectively, $5\textrm{D}_{3/2}(F=3)$ (Fig.\,\ref{fig6},
top) and $5\textrm{D}_{3/2}(F=2)$ (Fig.\,\ref{fig6}, bottom).
Since the probe beam radiation always contains a fraction of
linearly polarized light if projected on the $z$-axis (tilted by
11$^\circ$ with respect to $\bm{k}_\textrm{probe}$), all mentioned
hyperfine components (except $5\textrm{D}_{3/2}(F=3)$) contain two
different magnetic sublevels, as follows from Fig.\,\ref{fig3}.
After projecting on the $z$-axis,  the circularly polarized pump
beam will consist of 96\%  circular polarization (the same sign)
and 4\%  linear polarization.

\begin{figure}[t!]
\includegraphics[width= 0.9\linewidth]{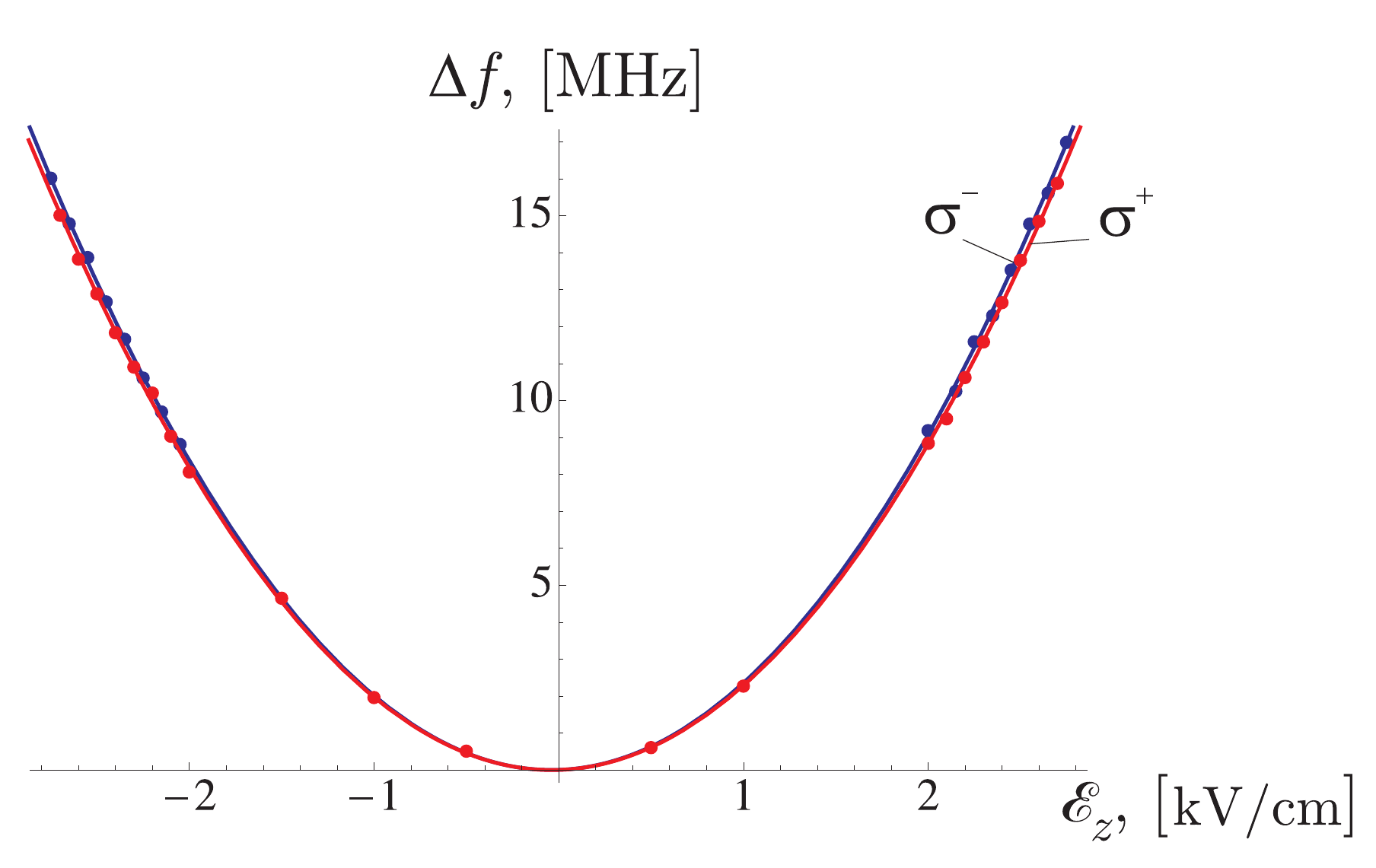}
\caption{ (Color online) Frequency shift of transitions to the
$5\textrm{D}_{5/2}(F=4)$ and $ 5\textrm{D}_{5/2}(F=3)$ hyperfine
sublelevs at different values of electric field $\mathcal{E}_z$.
The polarization of the probe beam along $\bm{k}_\textrm{probe}$
was $\sigma^+$ and $\sigma^-$, respectively. A parabolic fit
according to (\ref{eq4}) is used.
 } \label{fig7}
\end{figure}

Fig.\,\ref{fig6} shows an example of the data for the
$5\textrm{D}_{5/2}(F=4)$ and $ 5\textrm{D}_{5/2}(F=3)$ hyperfine
sublelevs obtained at different electric field strengths. The
difference in the two curves is due to the tensor polarizability.
Fitting the data with parabolic dependence, $\Delta f=p\mathcal{E}^2$,
according to (\ref{eq3},\ref{eq4}) allows us to derive the sensitivity $p$ of
each  transition to the  electric field. The results are:
\begin{eqnarray}\label{eq7}
p(5\textrm{D}_{5/2}(F=4),\,\sigma^+)&=&2.014(8)\,\textrm{MHz}/(\textrm{kV/cm})^2 \,,\\
\nonumber
p(5\textrm{D}_{5/2}(F=3),\,\sigma^+)&=&2.087(8)\,\textrm{MHz}/(\textrm{kV/cm})^2\,,\\
\nonumber
p(5\textrm{D}_{3/2}(F=3),\,\sigma^+)&=&2.066(8)\textrm{MHz}/(\textrm{kV/cm})^2\,, \\
\nonumber
p(5\textrm{D}_{3/2}(F=2),\,\sigma^-)&=&2.158(9)\textrm{MHz}/(\textrm{kV/cm})^2\,,\\
\nonumber
\end{eqnarray}
where the entry in the parentheses after the level notation 
denotes the polarization of the probe beam.

The spectrum shown in Fig.\,\ref{fig6}\,(top) contains only one
hyperfine component, $5\textrm{D}_{3/2}(F=3)$.  The
$5\textrm{D}_{3/2}(F=2)$  one  is not distinguishable from the
noise. This component can be excited by the linearly polarized
fraction of the probe beam from the
$5\textrm{P}_{3/2}(F=3,\,m_F=+2)$ sublevel. Knowing the fraction
of linear polarization in our probe beam, we can set a limit
to the $5\textrm{P}_{3/2}(F=3,\,m_F=+2)$ sublevel population (see
Table\,\ref{table2}). Table \ref{table2} also shows the relative
probabilities to excite different magnetic sublevels of the final
$5\textrm{D}$ state, which are calculated using (\ref{eq4}).

\begin{table} [t!]\caption{ The upper part of the table
shows the quality of the optical pumping derived from Fig.\,\ref{fig6}\,(top).
The lower part presents relative probabilities to excite relevant
hyperfine and magnetic sublevels of the $5\textrm{D}$ level at
different polarizations of the probe beam.
 }\label{table2}
\begin{ruledtabular}
\begin{tabular} {c|c|c} 
 initial state& $F,\,m_F$ & population\\
  \hline
$5\textrm{P}_{3/2}$&3,\ 3&$>0.97$\\
$5\textrm{P}_{3/2}$&3,\ 2&$<0.03$\\
\end{tabular}
\end{ruledtabular}
\vspace{0.01\textheight}
\begin{ruledtabular}
\begin{tabular} {c|c|c|c} 
 final state& $F,\,m_F$ & probe polarization & excitation \\
 &  & (with respect to $\bm{k}_\textrm{probe}$) & probability\\
  \hline
$5\textrm{D}_{5/2}$&4,\ 4&$\sigma^+$& 0.64\\
$5\textrm{D}_{5/2}$&4,\ 3&$\sigma^+$&0.01\\
$5\textrm{D}_{5/2}$&3,\ 3&$\sigma^-$& 0.004\\
$5\textrm{D}_{5/2}$&3,\ 2&$\sigma^-$&0.032\\
\hline
$5\textrm{D}_{3/2}$&3,\ 3&$\sigma^+$&0.024\\
$5\textrm{D}_{3/2}$&3,\ 2&$\sigma^-$& 0.192\\
$5\textrm{D}_{3/2}$&2,\ 2&$\sigma^-$& 0.192\\

\end{tabular}
\end{ruledtabular}
\end{table}

Using the results (\ref{eq7}), Table\,\ref{table2}, and relation
(\ref{eq4}) we get a system of linear equations for different magnetic
and hyperfine sublevels from which we derive scalar and tensor
polarizabilities. Polarizabilities for the 5P level are taken from
\cite{Windholz}, their uncertainties negligibly contribute to our
error budget.

Although the Rb cloud resides in the minimum of the MOT magnetic
field, we expect a residual magnetic field on the order of 0.1\,G
due to the final cloud size and adjustment imperfections.

Switching off the magnetic field in the experiment takes about 500\,$\mu$s, which would drastically reduce the duty cycle and the count rate. Therefore we decided to take the magnetic field into account instead of switching it off.

The influence of the magnetic field can be considered as additional degradation of the efficiency of the optical pumping. We found that magnetic field reduces relative population of the $5P_{3/2} (F=3, m=3)$ level in our experiment from 98\% to 97\%. More information about optical pumping in the magnetic field can be found in Appendix I. 

To verify that the field
does not significantly influence  the results of our measurements, we
measured the dependencies similar to that shown in
Fig.\,\ref{fig7} at  axial magnetic field gradients of
10\,G/cm and 5\,G/cm. We did not observe any significant difference within
the statistical uncertainty. Since our regular measurements were
performed at a gradient of 10\,G/cm, we conservatively add a systematic
uncertainty to the $p$ values (\ref{eq7}) of 0.1\% due to the
influence of the magnetic field.

The different fitting procedures  (a,b,c) described above  influence the
 results at the level of 0.03\,\%, so we add
this value to the uncertainty of coefficients $p$ coming from the
line shape model.

Calculation shows that the residual population of the
 5P$_{3/2}(F=3,\,m_F=2)$ level (see
Table\,\ref{table2}) can influence the coefficients $p$ on the level
of 0.07\%. Imperfection of the probe beam polarization and error in
determination of the angle between the $z$-axis and
$\bm{k}_\textrm{probe}$ also result in an uncertainty of 0.1\,\%.

\begin{table} [t!]\caption{ Uncertainty budget for coefficients $p$
describing the quadratic dependency on electric field (\ref{eq4}),
(\ref{eq7}). Uncertainty of electric field determination is
multiplied by 2 and included in this Table.}\label{table3}
\begin{ruledtabular}
\begin{tabular} {l|c} 
Effect& Uncertainty, \%\\
\hline
Statistical uncertainty& 0.2\\
Electric field determination&0.3\\
Residual magnetic field&0.1\\
Line shape model&0.03\\
Optical pumping&0.07\\
Probe beam polarization&0.1\\
AC Stark shift&0.1\\
{\bf Sum}&0.41\\
\end{tabular}
\end{ruledtabular}
\end{table}

The ac Stark shift of the 5P$_{3/2}$ level caused by a strong pump
beam may influence the result of the measurement if the pump and
probe pulses overlap. Experimental study of this effect shows that for the time delay used in the experiment (50\,ns), the residual pump beam perturbs the result
on the level of 0.1\%.

All mentioned uncertainties, including the uncertainty of electric
field determination, are summarized in Table\,\ref{table3}. Adding up all contributing uncertainties quadratically, we get 0.4\% for the  coefficient $p$. This uncertainty directly converts in the uncertainty of scalar and tensor
polarizability uncertainties as follows from (\ref{eq4}). Since
the main contribution to $p$ values comes from the scalar
polarizability $\alpha_S$, its relative uncertainty is similar to
the relative uncertainty of $p$. Tensor polarizability $\alpha_T$
is more than ten times smaller compared to $\alpha_S$, which means
that its relative uncertainty is  larger.

Scalar and tensor polarizabilities for the 5$D_{3/2,\,5/2}$ levels
measured in our experiment and corresponding uncertainties are
summarized in Table\,\ref{table4}.

\begin{table} [!t]\caption{Results for the tensor and scalar polarizability
measurements of the 5$D_{3/2,\,5/2}$ levels in Rb-87 in atomic
units. Here $\sigma$ is a standard deviation.}\label{table4}
\begin{ruledtabular}
\begin{tabular}{ |l|c|c| } 
 \hline
 polarizability & value [atomic un.] & uncertainty, $\sigma$  \\
 \hline
 $\alpha_S(5\textrm{D}_{3/2})$ & 18 400 & 75\\
 \hline
 $\alpha_T(5\textrm{D}_{3/2})$ & $-$750 & 30\\
 \hline
 $\alpha_S(5\textrm{D}_{5/2})$ & 18 600 & 76\\
 \hline
 $\alpha_T(5\textrm{D}_{5/2})$ & $-$1440 & 60 \\
 \hline
\end{tabular}
\end{ruledtabular}
\end{table}

In conclusion, we determined the scalar and tensor polarizabilities
of the 5D level in Rb. Using laser cooled atoms placed in a
constant electric field and two-step laser excitation we
demonstrated a relative uncertainty of 0.4\% for the scalar
polarizability and 4\% for the tensor polarizability. The demonstrated uncertainty
for the scalar polarizability  is comparable to accurate
measurements  in ground state alkali atoms. Our result is close to
the theoretical prediction \cite{Ovsiannikov} where the model
potential approach was implemented.

We are grateful to I.\,Veinstein and V.\,Ovsiannikov for
discussions and acknowledge support from RFBR grants  \#12-02-00867-a and \#11-02-00987-a.

\section{Appendix I: Optical pumping in the magnetic field}

\begin{figure}[t!]
\includegraphics[width= 0.9\linewidth]{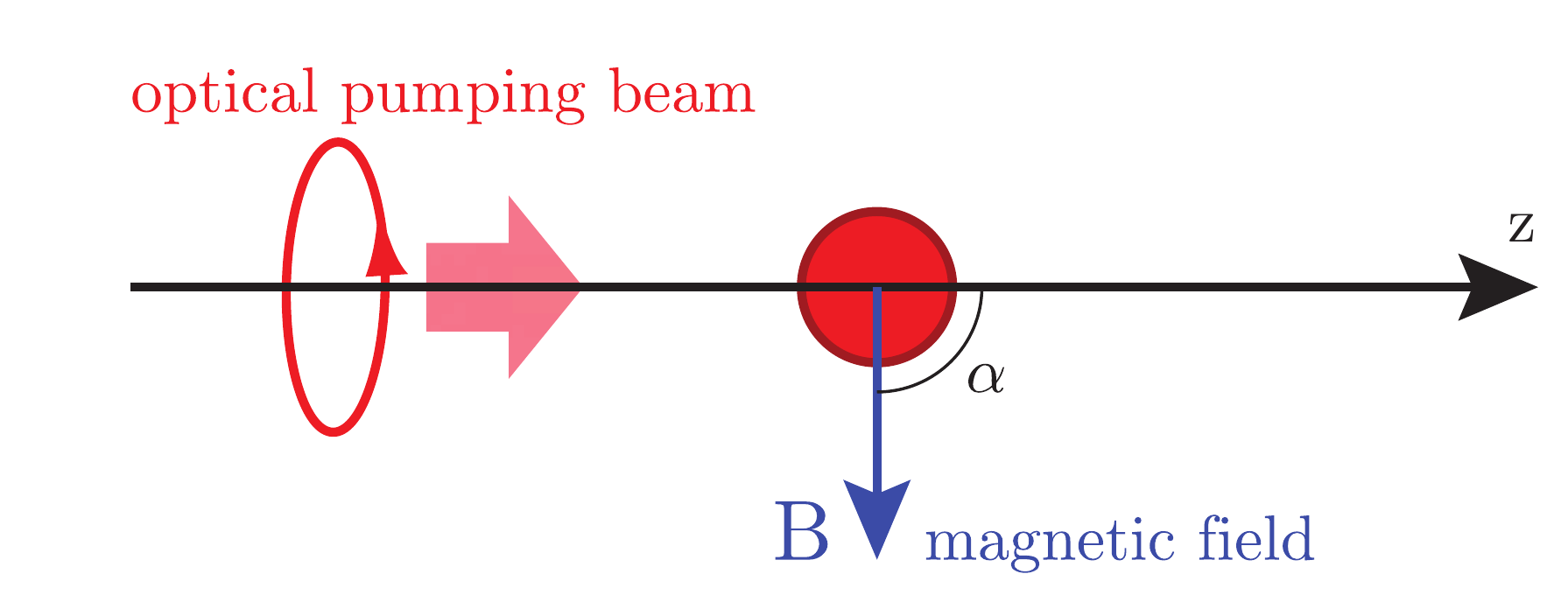}
\caption{ (Color online) An atom in a pumping beam and an external magnetic field. The pumping beam is propogating along the $z$ axis, which is selected as a quantization axis in the initial basis. Maximum influence of the magnetic field on the optical pumping process will take place when magnetic field is perpendicular to the $z$ axis. 
 } \label{fig8}
\end{figure}

Consider an atom placed in the pumping beam propagating along the $z$ axis (Fig.\,\ref{fig8}). A magnetic field directed at some angle to the $z$ axis will cause precession of the magnetic moment of the atom and therefore changes in the populations of magnetic sublevels will occur. The maximum influence of the magnetic field will take place when it is perpendicular to the $z$ axis.

Changes in populations of magnetic sublevels due to optical pumping can be found
by solving the master equation for the density matrix. Influence of the magnetic field can be calculated by projecting  the initial wavefunction onto the basis with quantization axis along the direction of the magnetic field.

The initial wavefunction is $|\psi\rangle=\sum{c_m |m\rangle}$ where $|c_m|^2$
describes populations of the magnetic sublevels with projection of the magnetic moment onto the $z$ axis equal to $m$. In the new basis this wavefunction will have a form $|\psi\rangle=\sum{c_{m_B}|m_B\rangle}$
where $|m_B\rangle$ is a state with projection of the moment onto the direction
of the magnetic field equal to $m_B$ and the coefficients $c_{m_B}$ can be calculated
according to the equation:

\begin{equation}\label{eqP}
c_{m_B}=\sum_m{c_m\int{Y^F_m(\theta,\phi)Y^{F}_{m_B}(\theta',\phi')^*\sin{\theta}
d\theta d\phi}}.
\end{equation}

Here the angles $\theta'$ and $\phi'$ can be expressed over
$\theta$,$\phi$ and $\alpha$, where $\alpha$ is an angle between the original
quantization axis and the magnetic field. If the magnetic field is perpendicular to the $z$ axis we have $\alpha=\pi/2$ and $\theta'=\arccos(\sin \theta\,\cos\phi)$, $\phi'=\arctan(\sin\phi\,\tan\theta)$. Taking into account
evolution of the wavefunction in time we obtain: 

\begin{equation}
|\psi\rangle=\sum{c_{m_B}|m_B\rangle e^{-i \frac{\mu_B g_F}{\hbar}
 m_B B dt}},
\end{equation}
 
\noindent where $g_F$ is the hyperfine Lande g-factor. Returning back to
the initial basis, as in (\ref{eqP}), we will obtain the new populations of the magnetic sublevels.

\begin{figure}[t!]
\includegraphics[width= 0.9\linewidth]{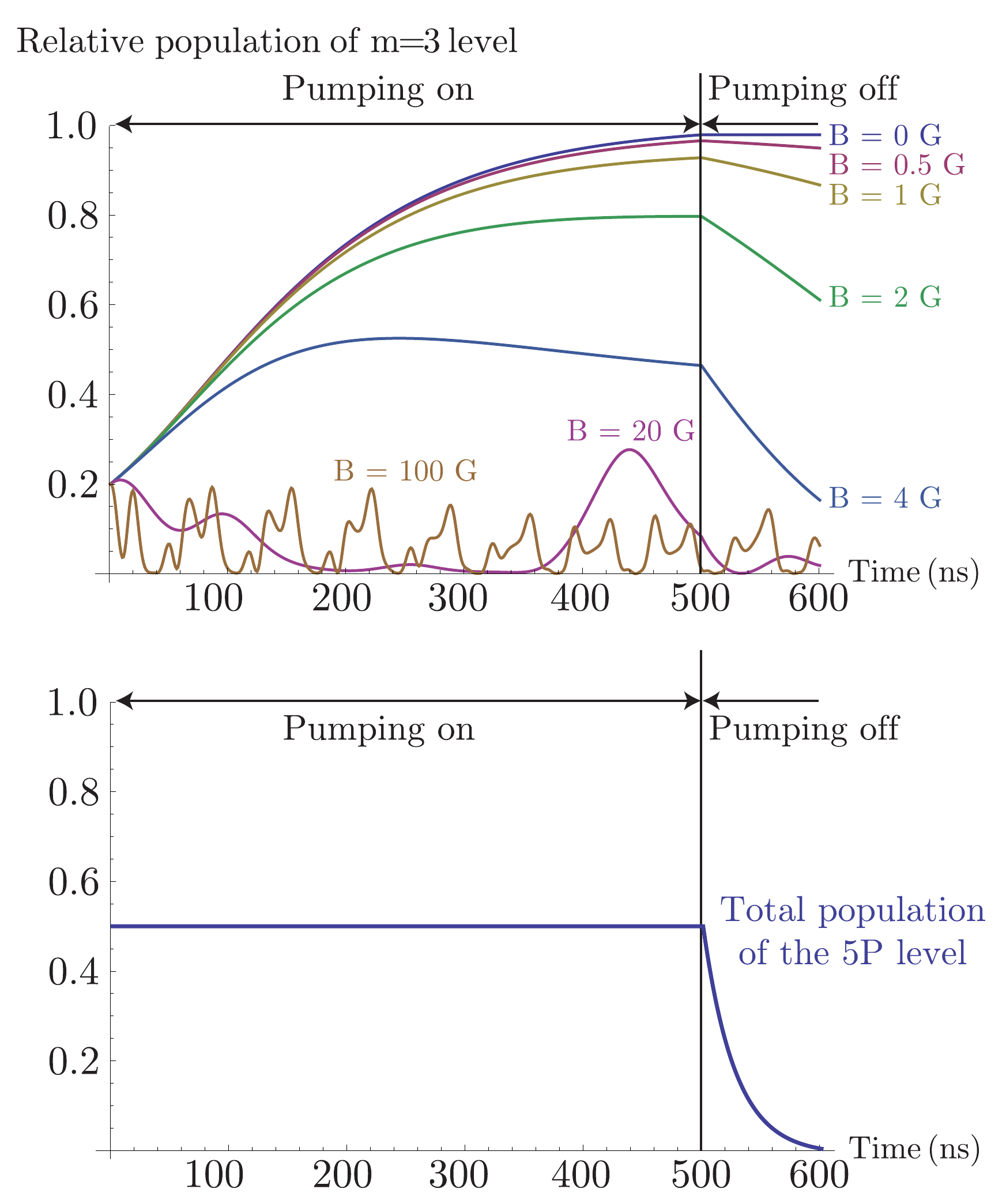}
\caption{ (Color online) Top: numerical calculations of the population of
the magnetic level of $5P_{3/2}$ with $m=3$ during
the optical pumping and after that. The population of the magnetic
sublevel is normalized to the total population on the $5P_{3/2}$ level. Bottom: total population of the $5P_{3/2}$ level.
 } \label{fig9}
\end{figure}

We accomplished numerical calculations of the population dynamics in the magnetic field perpendicular to the $z$ axis with 500\,ns of optical pumping and further 100\,ns evolution of the atom in the magnetic field without pumping. The results of the  simulations for different values of the applied magnetic field  are shown in Fig.\,\ref{fig9}. We can see that optical pumping remains efficient with magnetic fields up to 2\,G. Larger fields leads to rapid mixing of states even during the pumping. Imperfection in the optical pumping in the experiment was about 3\%,  corresponding to a magnetic field less than 0.3 G. 

\end{document}